\documentstyle[prd,aps,floats]{revtex}
\begin{document}
\draft

\input epsf \renewcommand{\topfraction}{0.8}
\twocolumn[\hsize\textwidth\columnwidth\hsize\csname
@twocolumnfalse\endcsname

\hfill HIP-2002-18/TH

\title{Reheating as a surface effect}
\author{Kari Enqvist~$^{a,b}$, Shinta Kasuya~$^{b}$,
and Anupam Mazumdar~$^{c}$}
\address{$^{a}$ Department of Physical Sciences,
P. O. Box 64, FIN-00014, University of Helsinki, Finland.\\
$^{b}$ Helsinki Institute of Physics, P. O. Box 64, FIN-00014,
University of Helsinki, Finland.\\
$^{c}$ The Abdus Salam International Centre for Theoretical Physics,
I-34100, Trieste, Italy.}
\maketitle

\begin{abstract}
We describe a new mechanism for reheating the Universe
 through evaporation of a surface charge of a fragmented
inflaton condensate.  We show that
for a range of Yukawa coupling of the inflaton to the matter sector
 evaporation  gives rise to a much smaller
reheat temperature compared to the standard perturbative decay.
As a consequence, reheating through a surface effect could solve the gravitino
and moduli over production problem in inflationary models without fine
tuning the Yukawa sector.
\end{abstract}

\vskip2pc]

Inflation  gives naturally the initial condition for the hot big bang
cosmology by reheating the Universe through a conversion of the energy
density stored in the inflaton field into radiation of elementary
particles \cite{linde1}. However, as it is well known, the entropy thus
dumped into the Universe may pose a problem for big bang nucleosynthesis.
This constraints the number densities of light quanta (or massive quanta which
later decay) at the time of reheating, which is given by \cite{ellis}
${n}/{s} \approx \eta (T_{\rm R}/10^{11})~{\rm GeV}$,
where $n$ is number density, $s$ is the entropy density of the thermal
bath with a reheat temperature $T_{\rm R}$, and $\eta\approx 10^{-11}$
is the matter-to-entropy ratio at the time of big bang nucleosynthesis.
For a successful big bang nucleosynthesis one requires
$n/s \leq 10^{-13}$, which translates into $T_{\rm R} \leq 10^9~{\rm GeV}$
\cite{subir}. Such a low reheat temperature is a challenge to inflation
models as well as a perennial problem. For example, in models involving
supersymmetry and superstrings one is often faced with an overproduction
of gravitinos and moduli unless the reheat temperature can somehow be
made low \cite{ellis,subir}.

Reheating is based on the decay of the inflaton field
while it oscillates about its minimum \cite{kolbturner}.
The amplitude of the oscillations gradually decreases
as the Universe expands, and the energy density redshifts as
$\rho\propto a^{-3}$, where $a$ is the scale factor of the Universe.
There exists two conceptually different realizations for the decay:
direct (and slow) decay of the inflaton quanta, and a resonant
conversion of the inflaton field to decay products by a process
called preheating \cite{traschen}.

The decay lifetime of the inflaton can be computed in the limit $m\gg H$,
where $m$ is the inflaton mass and $H$ is the expansion of the Universe.
In particular decay of an inflaton to a pair of fermions with an interaction
${\cal L}_{int}=h\phi\bar\psi\psi$, when $m\gg m_{\psi}$, is given by
\begin{equation}
\label{gammadir}
\Gamma(\phi\rightarrow \bar\psi \psi)=\frac{h^2m}{8\pi}\,,
\end{equation}
where $h$ is the Yukawa coupling smaller than one. In order to estimate
the reheat temperature one equates the decaying
inflaton energy density to the radiation energy density;
$\rho_{r}=(\pi^2/30)g_{\ast}T_{R}^4$, at the moment when $\Gamma =H$.
Assuming that thermalization occurs quickly, one obtains a reheat temperature
\begin{equation}
\label{reh2}
T_{R}=\left(\frac{90}{8\pi^3g_{\ast}}\right)^{1/4}\sqrt{\Gamma M_{\rm P}}
\sim 0.1\sqrt{\Gamma M_{\rm P}}\,.
\end{equation}
where $g_{\ast}\sim {\cal O}(100)$ is the relativistic degrees of freedom
at the time of reheating, and $M_{\rm P}=1.2\times 10^{19}$ GeV.

Preheating requires specific conditions which may or may not occur.
However, the important point is that the standard reheating phenomena
is a volume effect. If instead of the inflaton decaying inside some
fixed volume it were to evaporate through the surface, the decay rate
would naturally be suppressed by the ratio
\begin{equation}
\label{con}
\frac{{\it area}}{{\it volume}} \propto R^{-1}\,,
\end{equation}
where $R$ is the effective size of an object whose surface is evaporating.
The larger the radius, the smaller is the evaporation rate, and therefore,
the smaller is the reheat temperature. Reheating through a surface effect
would thus be a natural way to achieve low reheat temperatures.

Note that such a phenomenon takes place in a
self gravitating system such as a black-hole. Indeed, it has been
claimed that in case of a self-gravitating object there holds a
holographic principle according to which all the degrees of freedom
lie on the surface \cite{thooft}. A black-hole with a mass $M_{bh}$
evaporates from its gravitational radius $r_{g}=2M_{bh}/M_{\rm P}^2$ as
a black body with temperature inversely proportional to the radius of a black
hole $T_{bh}=(M_{\rm P}^2/8\pi M_{bh})$ \cite{hawking75a}. However,
because of the small density contrast in the early Universe implied by
the anisotropies of the cosmic microwave background, a sufficient number
of primordial black holes might be very difficult to form.

However, a non-topological defect, which might encompass
an inflaton condensate within a sub-horizon radius,  would typically
also decay through its surface. An example is the $Q$-ball \cite{coleman}
that can exist in theories with scalar fields carrying a conserved global
$U(1)$-charge. The $Q$-balls are long-lived non-topological solitonic
ground state solutions for a fixed charge $Q$. This means that the energy
of a $Q$-ball configuration is less than a
collection of free scalars carrying an equivalent charge. The $Q$-ball field
configuration is given by $\phi(x,t)= e^{i\omega t}\varphi(x)$, its
energy and charge are given by
\begin{eqnarray}
\label{energy}
E&=&\int d^3x\left[\frac{1}{2}\omega^2\varphi^2+\frac{1}{2}
|\nabla \phi^2|+V(\phi)\right]\,, \nonumber \\
Q &=&\omega\int \varphi^2 d^3x\,.
\end{eqnarray}
The potential $V(\phi)$ has a global minimum at the origin and it is invariant
under global $U(1)$-transformation. Depending on the slope of the potential
it is possible to have energetically stable $Q$-ball, for which we require
$E\leq \mu Q$, where $\mu$ is given by
\begin{equation}
\label{mu}
\mu ={\rm min}\left(\sqrt{\frac{2 V(\phi)}{\phi^2}}\right)\,.
\end{equation}
In order to form $Q$-balls a negative pressure is required for the scalar
condensate to fragment \cite{kusenko,em}. Several analytical and numerical
studies have verified this in the context of the MSSM flat direction
Affleck-Dine condensate \cite{km}.

An example of a potential which can lead to a $Q$-ball formation is
given by
\begin{equation}
\label{qpot}
V(\varphi)=m^2\left(1+|K|\log\left[\frac{\varphi^2}{M_{\rm P}^2}\right]\right)
\varphi^2\,,
\end{equation}
The one-loop logarithmic running of $\varphi$ appears due to the presence
of finite Yukawa couplings to bosons and fermions, or, if $\varphi$ has
some gauge interactions. Such a correction to the potential appears in
supersymmetry and in particular if $\varphi$ has gauge interactions
then $K<0$ \cite{em}. In non-supersymmetric theories it is possible
to obtain $K<0$, provided the Yukawa coupling to the fermions dominate,
or if there are more fermionic loops than bosonic ones.

Let us now assume that our inflaton sector has a global $U(1)$ symmetry
with a running mass governed by Eq.~(\ref{qpot}). Although the inflaton
is often taken to be a real, there seems to be no reason why it should not
be a complex field (and perhaps this would be even more natural).
Then chaotic inflation occurs for initial  values $\varphi > M_{\rm P}$
\cite{linde1}. The running mass inflation has been a topic of recent
investigation and interesting details can be found in \cite{liddle-lyth00}.

We note that right after the end of inflation, when the inflaton condensate
starts to oscillate at $H\approx m$, the condensate can fragment due to the
perturbations in $\phi$ field and eventually form a $Q$-ball.
In this case $Q$-balls can form due to
the oscillations in $\phi$ field \cite{km,kasuya}.

The oscillating inflaton field will fragment into $Q$-balls if $|K|<0$.
To see this, the potential Eq.~(\ref{qpot}) can be
expanded in a form $V(\varphi)\propto \varphi^{2+2|K|}$. The equation of
state for such a potential is given by $p=-(K/2)\rho$ \cite{turner}. Hence,
oscillating inflaton field does not behave effectively as pressureless matter,
as it is usually assumed, but feels a negative pressure.
A perturbation in the inflaton field will then obey \cite{em}
\begin{equation}
\delta \ddot\varphi_{\rm k}=-\frac{K{\rm k}^2}{2}\delta\varphi_{\rm k}\,.
\end{equation}
Therefore, the fluctuation of the field grows on scale
${\rm k}=2\pi/\lambda$ in time
\begin{equation}
\label{pert1}
\delta \varphi_{\rm k}=\delta \varphi_{i~{\rm k}}\exp\left(\sqrt{\frac{
|K|{\rm k}^2}{2}}~t\right)\,,
\end{equation}
where $t=0$ corresponds to the beginning of a coherent oscillations of the
inflaton field, and $\delta\phi_{i}$ is the initial perturbation in the
field when $H\sim m$. Here, obviously we have neglected the effect of
expansion, but we keep in mind that a physical momentum redsifts during the
expansion of the Universe, and as long as the wavelength of the excited
mode is well within the Hubble length it is possible to follow the
fragmentation of the inflaton condensate. Note that perturbations of
wavelength $\lambda$ take a finite time to grow non-linear
$t\sim ({1}/{2\pi})\left({2}/{|K|}\right)^{1/2}\lambda$.
 This may be taken as an estimate for the time it takes to fragment
the inflaton condensate and to form lumps of charged $Q$-balls
\cite{km}.

A potential of the form Eq.~(\ref{qpot}) generically gives rise to a
thick-wall $Q$-ball with a Gaussian profile
$\varphi(r)\propto \exp(-|K|m^2r^2/2)$. The size
of such a $Q$-ball can be estimated by \cite{em}
\begin{equation}
\label{rad}
R\equiv |K|^{-1/2}m^{-1}\,,
\end{equation}
Therefore, a $Q$-ball of size $R$ forms when
\begin{equation}
H_{f} \propto \frac{1}{t} \sim {\cal O}(1)|K|m\,.
\end{equation}
The exact proportionality depends on details which we are neglecting
at the moment for an order of magnitude estimate.

The total charge accumulated by a $Q$-ball will therefore depend on
$Q=2\omega \varphi^2_{0}V=(8\pi/3)m \varphi^2_{0}R^3$, where
$\varphi_{0}$ is the inflaton field value at which the $Q$-ball forms. Since
$|K|<1$, we can approximate the decay in the amplitude of the oscillations
by $\varphi_{0} \sim \varphi_{i}(H_{f}/H_{i})$ similar to a matter dominated
era, where $\varphi_{i} \sim M_{\rm P}$, denotes the end of inflation
in chaotic model, and $H_{i}\sim m$ when the oscillations begin.
Therefore, the total charge of a $Q$-ball can be given by
\begin{eqnarray}
Q \sim \frac{3}{8\pi}\left(\frac{1}{|K|^{1/2}m}\right)^3
|K|^2 mM_{\rm P}^2 = \frac{3}{8\pi}|K|^{1/2}\left(\frac{M_{\rm P}}{m}
\right)^2\,.
\end{eqnarray}
Note, that the charge is fixed by the ratio of $m/M_{\rm P}$ determined
by the anisotropies seen in the cosmic micro wave background radiation
\cite{liddle-lyth00}
\begin{equation}
\frac{\Delta \rho}{\rho} \sim \frac{m}{M_{\rm P}}\sim 10^{-5}\,,
\end{equation}
which results in $m\sim 10^{13}$~GeV.

The evaporation of a $Q$-ball through the decay of a charged massless fermion
has been studied in \cite{cohen,multamaki}. Analytical estimates in the
thin-wall case have yielded an upper limit on the evaporation
rate, which is saturated even in thick-wall case for
$h\varphi_{0}\geq \omega$, where $h$ is the Yukawa coupling \cite{cohen}.
The evaporation rate is bounded by
\begin{equation}
\label{evap}
\frac{dQ}{dt} \leq \frac{\omega^3 A}{192\pi^2}\,,
\end{equation}
where in our case $\omega=m$. The above estimation is valid for
strong Yukawa coupling. In Eq. (\ref{evap}), $A=4\pi R^2$
denote the surface area of a $Q$-ball, and the above expression simplifies to
\begin{equation}
\label{evap1}
\frac{dQ}{dt} \leq \frac{1}{48\pi}\frac{m}{|K|}\,.
\end{equation}
Therefore, we can estimate the decay rate of a $Q$-ball in the process of
reheating the Universe
\begin{equation}
\label{evap2}
\Gamma =\frac{1}{Q}\frac{dQ}{dt}\approx \frac{m}{18 |K|^{3/2}}\left(
\frac{m}{M_{\rm P}}\right)^2 \,.
\end{equation}
In general $K$ and $h$ are not independent quantities but
are related to each other by $K \sim C h^2/16\pi^2$, where $C$ is effective
number of bosonic and fermionic loops.
Even though we are in the strong coupling limit, the decay rate mimics that
of a Planck suppressed interaction of the inflaton made $Q$-ball to the
matter fields. The estimated reheat temperature is then given by
\begin{equation}
\label{fnl1}
T_{R} \leq  \frac{1}{|K|^{3/4}}10^{8}~{\rm GeV} \approx 5C^{-3/4} h^{-3/2}\times
10^{9}~{\rm GeV}\,.
\end{equation}
If the inflaton sector does not belong to a hidden sector then it
is very natural that inflaton coupling to other matter fields is
sufficiently large, i.e. $1\gtrsim h\gtrsim (m/M_{\rm P})$. As an example, we may
consider $C \sim 10$, and $h \sim 1$, we obtain $|K| \sim 0.1$. The
resulting reheat temperature is then $T_{R} \approx 5\times 10^8 ~{\rm GeV}$.
We notice that the dependence of the reheat temperature on $|K|$ is
in general rather weak. For such a low reheat temperature gravitinos and
moduli can never be over produced from a thermal bath \cite{ellis}.
Therefore, reheating through a surface of a $Q$-ball solves the gravitino
and moduli problem without invoking very small Yukawa coupling. This is
the main application of reheating through a surface.

We notice that for couplings $1\gtrsim h\gtrsim {\cal O}(m/M_{\rm P})$,
Eq.~(\ref{fnl1}) predicts in this particular model much lower reheat temperature than
direct decay of inflaton, which from Eq.~(\ref{gammadir}), yields
\begin{equation}
\label{fnl2}
T_{R}\sim 2\times 10^{14}h ~{\rm GeV}\,.
\end{equation}

In the weak Yukawa coupling case when $h\varphi_{0}\ll m$
(so that $h\ll m/M_{\rm P})$, the evaporation
rate of a $Q$-ball \cite{cohen}, and subsequently the decay rate are bounded by
\begin{equation}
\label{sh}
\frac{dQ}{dt} \leq \frac{1}{16}\frac{h\varphi_{0}}{|K|}\,,~~~
\Gamma \leq \frac{h\pi}{6|K|^{1/2}}\frac{m^2}{M_{\rm P}}\,,
\end{equation}
and the limit on the reheat temperature reads now
\begin{equation}
T_{R}\leq 0.1\left(\frac{h}{|K|^{1/2}}\right)^{1/2}m \quad \sim \quad
0.3 C^{-1/4}m \,.
\end{equation}
Note that the dependence on $h$ is cancelled out.
Thus for a sufficiently small $h\ll (m/M_{\rm P})\sim 10^{-5}$ the
reheat temperature saturates and would be larger than through the
perturbative decay of inflaton. This result may appear counter-intuitive
as it suggests that the evaporation rate is faster than the perturbative
decay rate for small $h$. The point is that a fermion gets an effective
mass $m_{\psi} \sim h\varphi_0$ inside a $Q$-ball. The corresponding
Compton wavelength $m_{\psi}^{-1}$ then actually becomes larger than
the size of a $Q$-ball for very small $h$. In fact the ratio between
the evaporation and perturbative decay rate goes as
$\sim m_{\psi}^{-1}/R > 1$, and therefore the evaporation through
a surface effect given by Eq.~(\ref{sh}) cannot be trusted any
longer. This is an example where a simple relationship given by
Eq.~(\ref{con}) does not hold.

We have pointed out a new possibility of reheating the Universe through an
evaporation of a surface. We have shown that without invoking very low
fine tuned Yukawa coupling, the evaporation of a surface charge from the
inflaton condensate inevitably gives much smaller reheat temperature
compared to the standard perturbative decay of inflaton. This new mechanism
could be perhaps the simplest option to solve the gravitino, polonyi
and dilaton problems in general \cite{subir}.

Even though we have considered $Q$-ball formation and reheating of the
Universe in a chaotic inflationary scenario, it is conceivable that
similar behaviour could arise in many other inflaton models
with an effective $U(1)$ global charge. It is quite possible that
the inflaton condensate fragmentation and $Q$-ball formation can occur
even if $U(1)$ would be slightly broken. Hence, in addition to the
standard reheating, which is a volume effect, for a range of parameters
the Universe may be reheated in a novel way, through surface evaporation
of a $Q$-ball.


\vskip10pt
A.M. is thankful to Rouzbeh Allahverdi, and Altug Ozpineci for helpful
discussion, and S.K. is thankful to M. Kawasaki for a useful discussion.
A.M. acknowledges the support of {\it The Early Universe network}
HPRN-CT-2000-00152, and a kind hospitality of the Helsinki Institute of
Physics where this work has been carried out.



\end{document}